\documentclass[aps,prd,twocolumn,superscriptaddress]{revtex4}
\usepackage{graphicx}
\usepackage{dcolumn}
\usepackage{hyperref}

\def\beq{\begin{equation}}
\def\eeq{\end{equation}}
\def\bea{\begin{eqnarray}}
\def\eea{\end{eqnarray}}

\def\v{\varphi}

\def\sgn{{\rm sgn}}
\def\bwt{\begin{widetext}}
\def\ewt{\end{widetext}}

\nocite{*}

\usepackage{hyperref}

\begin{document}

\title{Cuscuton Cosmology: Dark Energy meets Modified Gravity}
\author{Niayesh Afshordi}\email{nafshordi@cfa.harvard.edu}\affiliation{Institute for
  Theory and Computation,
Harvard-Smithsonian Center for Astrophysics, MS-51, 60 Garden
Street, Cambridge, MA 02138, USA}
\author{Daniel J.H. Chung}\email{Danielchung@wisc.edu}
\affiliation{Department of Physics, University of Wisconsin,
  Madison, WI 53706, USA}
  \author{Michael Doran}\email{M.Doran@thphys.uni-heidelberg.de}
\affiliation{Institut fur Theoretische Physik, Philosophenweg 16,
69120 Heidelberg, Germany}
\author{Ghazal Geshnizjani}\email{ghazal@physics.wisc.edu}
\affiliation{Department of Physics, University of Wisconsin,
  Madison, WI 53706, USA}
\date{\today}
\preprint{astro-ph/yymmnnn}
\begin{abstract}
In a companion paper \cite{companion2}, we have introduced a model of
scalar field dark energy, {\it Cuscuton}, which can be realized as the
incompressible (or infinite speed of sound) limit of a k-essence
fluid.  
In this paper, we study how {\it Cuscuton} modifies the constraint
sector of Einstein gravity.  In particular, we study {\it Cuscuton}
cosmology and show that even though {\it Cuscuton} can have an
arbitrary equation of state, or time dependence, and is thus
inhomogeneous, its perturbations do not introduce any additional
dynamical degree of freedom and only satisfy a constraint equation,
amounting to an effective modification of gravity on large scales.
Therefore, {\it Cuscuton} can be considered to be a minimal theory
of evolving dark energy, or a minimal modification of a cosmological
constant, as it has no internal dynamics. Moreover, this is the only
modification of Einstein gravity to our knowledge, that does not
introduce any additional degrees freedom (and is not conformally
equivalent to the Einstein gravity). We then study two simple {\it
Cuscuton} models, with quadratic and exponential potentials. The
quadratic model has the exact same expansion history as
$\Lambda$CDM, and yet contains an early dark energy component with
constant energy fraction, which is constrained to $\Omega_Q \lesssim
2\%$, mainly from WMAP Cosmic Microwave Background (CMB) and SDSS
Lyman-$\alpha$ forest observations. The exponential model has the
same expansion history as the DGP self-accelerating braneworld
model, but generates a much smaller integrated Sachs-Wolfe (ISW)
effect, and is thus consistent with the CMB observations. Finally,
we show that the evolution is local on super-horizon scales,
implying that there is no gross violation of causality, despite {\it
Cuscuton}'s infinite speed of sound.


\end{abstract}
\maketitle

\section{Introduction}

The nature of the current acceleration of cosmic expansion is among
the most outstanding puzzles in theoretical physics. Various
cosmological observations such as the dimming of distant supernovae
Ia \cite{Riess,Perlmutter:1998np}, anisotropies in the Cosmic
Microwave Background (CMB) \cite{Spergel:2006hy}, and the large
scale structure of the Universe (e.g. \cite{Seljak:2006bg}) can be
most easily explained by having an exotic {\it dark energy}
component with negative, nearly constant and uniform, pressure which
dominates the energy density of the Universe.

While the simplest model of vanilla dark energy, i.e. a cosmological
constant, remains consistent with all the present observations
\cite{Seljak:2006bg}, many models of non-minimal dark energy have been
developed in anticipation of any future failure of cosmological
constant in explaining the data. However, all the models that predict
an observable dark energy density evolution suffer from an extreme
fine tuning problem, requiring incredibly light masse scales ($\sim
10^{-33} {\rm eV}$) that are hard to protect from quantum corrections.

In a companion paper \cite{companion2}, we developed a new model of
field theoretical dark energy, {\it Cuscuton}, that while generally
non-uniform, lacks any dynamical degree of freedom. This protects
the theory from quantum corrections at low energies, and thus makes
it a perfect candidate for an evolving dark energy in the current
era. The name {\it Cuscuton} (pronounced {\tt
k\"{a}s-k\"{u}-t\"{a}n}), is derived from the Latin name for the
parasitic plant of dodder, Cuscuta.  Classically, it is a new kind of
constraint system, allowing a novel class of constrained dynamics.

The {\it Cuscuton} action for the scalar filed $\v$ can be written
as \beq S =\int
d^{4}x\sqrt{-g}[\mu^2\sqrt{|g^{\mu\nu}\partial_{\mu}\varphi\partial_{\nu}\varphi|}-V(\varphi)],\label{cuscutan}\eeq
where $\mu$ is an (arbitrarily defined) energy scale. One can show
that {\it Cuscuton} is an incompressible k-essence fluid
\cite{Armendariz-Picon:2000dh,Chiba:1999ka}, i.e. it has an infinite
speed of sound, raising questions about the causality of the theory.
However, in \cite{companion2}, we show that there is no breakdown of
causality, as the perturbations lack (symplectic) dynamics, and thus
do not carry information.

Given that {\it Cuscuton} acts as a constraint system, it is
interesting to ask how a minimal coupling of {\it Cuscuton} to
gravity will modify interaction of gravity with matter.  In
particular, as is well known, the scalar perturbations about the FRW
metric represent the general relativistic spacetime constraints for
matter coupled to gravity. A non-trivial choice of {\it Cuscuton}
potential will modify the constraint equations, leading to
observable consequences in cosmology. We aim to characterize these
observables associated with the scalar perturbations.

We derive general scalar perturbation equations in the presence of
{\it Cuscuton} and show analytically and numerically how the
constraints are modified.  We show that {\it Cuscuton} models can
have an expansion history identical to that of $\Lambda$CDM but with
different CMB and matter power spectra due to gravitational
potential evolution mimicking that of tracker models of
quintessence. We also find that {\it Cuscuton} can exactly replicate
the expansion history of the DGP self-accelerating cosmology, while
predicting similar small angle CMB and matter power spectra, at a
few percent level. The main difference between the DGP
self-accelerating cosmology and {\it Cuscuton} will show up due to
the anisotropic-stress induced boost in the Integrated Sachs-Wolfe
(ISW) effect for the DGP model. Finally, we briefly consider the
issue of causality in the scalar sector and find that a Yukawa-like
exponential fall-off of {\it Cuscuton} effects beyond the
$\sqrt{\dot{H}}$ scale protects the gravitational phenomenology from
gross anomalies in the long wavelength limit.  We show that the
usual Bardeen parameter remains approximately constant in the long
wavlength limit.

The paper is organized as follows: In Section \ref{back}, we compute
the background evolution, followed by linear perturbations in
Sections \ref{linear}. We then study how the matter power spectrum
and the ISW effect are affected by {\it Cuscuton} in Section
\ref{sec::linear}, and discuss current observational constraints.
Before presenting our conclusions in Section \ref{sec::conclusions},
we also discuss causality of linear {\it Cuscuton} perturbations in
Section \ref{sec::causality}.
Throughout this paper, we use the reduced Planck's constant
$M_{p}=(8\pi G_{N})^{-1/2}\approx2\times10^{18}$ GeV.

\section{Background Evolution}\label{back} Starting with the {\it Cuscuton}
action (Eq. \ref{cuscutan}) in a flat Friedmann-Robertson-Walker
(FRW) metric \beq ds^2= dt^2 - a(t)^2 dx^i dx^i, \eeq and a
homogenous field configuration, i.e. $\v = \v(t)$, the action takes
the form \beq S = \int a^3 dt \left[\mu^2|\dot{\v}| -
V(\varphi)\right]. \eeq Varying the action with respect to $\v$
yields the background field equation \beq (3\mu^2H) ~\sgn(\dot{\v})
+ V^{\prime}(\v) = 0.\label{0thfield}\eeq

 One implication of Eq.(\ref{0thfield}) is that $V(\v)$ always
decreases (increases) with time in an expanding (collapsing)
universe. Eq.(\ref{0thfield}), in combination with the Friedmann (or
$G^0_0$ Einstein) equation in a flat universe \beq H^2 =
\frac{\rho_{\rm tot}}{3 M^2_p},\label{f1}\eeq yields \beq
\left(M^2_p\over 3\mu^4\right) {V^{\prime}}^2(\v) - V(\v) = \rho_m,
\label{friedmann}\eeq where $\rho_m$ is the background density of
ordinary matter in the Universe. Notice that the {\it Cuscuton}
kinetic term does not contribute to its energy density.
\begin{figure}
\includegraphics[width=1.1\linewidth]{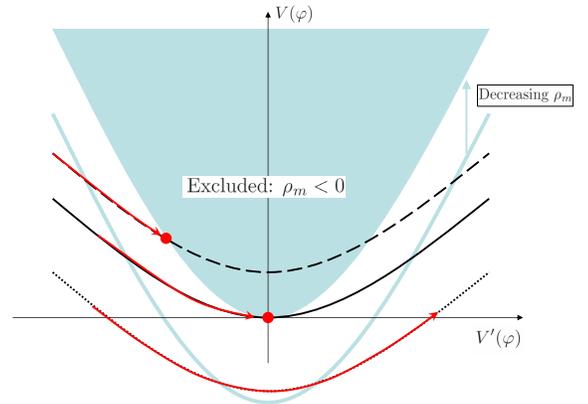}
\caption{\label{vvp}Different possible  {\it Cuscuton} potentials in
the $V-V^\prime$ phase space. The shaded region is excluded in a
flat FRW cosmology, as it requires negative matter density, $\rho_m$
(Eq. \ref{friedmann}). The dark solid, dashed and dotted lines are
quadratic potentials with zero, positive and negative bare
cosmological constants. The red (dark) arrows show the time
evolution of the field in a flat universe, and the points show the
asymptotic state of the field, if present. The blue (light) solid
line shows a constant $\rho_m
> 0$ contour. As we show in the text, in addition to avoiding the
excluded region, the potentials should be shallower than the
constant $\rho_m$ contour in the $V-V'$ plane.}
\end{figure}

In Fig.(\ref{vvp}), we show different possibilities for the {\it
Cuscuton} potential in the $V-V^\prime$ phase space. Eq.
(\ref{friedmann}) implies that for a positive matter density
($\rho_m >0$; weak energy condition), assuming a flat universe, we
should have \beq V(\v) < \left(M^2_p\over 3\mu^4\right)
{V^{\prime}}^2(\v). \eeq The excluded region is shown by the blue
shaded area in Fig. (\ref{vvp}), while the blue (light) line shows a
constant $\rho_m
>0$ contour.

The continuity equation for matter density $\dot{\rho}_m = -3H
(\rho_m + p_m)$, in combination with Eq.(\ref{0thfield}) and the
time derivative of Eq.(\ref{friedmann}), assuming the null energy
condition $\rho_m + p_m >0$,  gives: \beq
3\mu^2H|\dot{\v}|\left(\frac{2M^2_P}{3\mu^4}V^{\prime\prime}(\v)
-1\right) = -\dot{\rho}_m >0, \eeq which puts an additional
constraint on the potential: \beq V^{\prime\prime}(\v) = \frac{1}{2}
\frac{dV^{\prime 2}(\v)}{dV(\v)} > \frac{3\mu^4}{2M^2_p}. \eeq
Therefore, all the allowed potentials need to be shallower than the
constant density contours (blue/light line) in Fig.(\ref{vvp}).

Dark lines in Fig.(\ref{vvp}) show different quadratic potentials of
the form \beq V(\v)= V_0 + \frac{1}{2}m^2\v^2. \label{quadratic}
\eeq Interestingly -- as we  will show in Section \ref{sec::linear}
-- a quadratic potential leads to a tracking behavior. In this case,
the quadratic term of the potential always maintains a constant
fraction of the total  density of the Universe. The dark solid line
shows a quadratic potential with no bare cosmological constant
($V_0=0$). For a positive bare cosmological constant ($V_0>0$;
dashed line), it is interesting to notice that the potential never
reaches its minimum, and is stalled at the boundary of the $\rho_m <
0$ region (shaded area). Therefore, the effective value of the
cosmological constant is larger than its bare value. Finally, a
potential with a negative bare cosmological constant ($V_0<0$;
dotted line) results in a bounce, as $V^\prime$, and thus the
expansion rate go to zero, and changes sign subsequently.

Another interesting example is an exponential potential of the form
\beq V(\v) = V_0 \exp\left[-\left(\mu^2r_c\over
M^2_p\right)\v\right]. \eeq Direct substitution in Eqs.
(\ref{0thfield}-\ref{f1}) yields \beq H= \frac{1}{2r_c} +
\sqrt{\frac{1}{4r^2_c}+\frac{\rho_m}{3M^2_p}},\label{dgph}\eeq which
is exactly the same as the background dynamics in a (flat)
Dvali-Gabadadze-Poratti (DGP)\cite{Dvali:2000hr} 5D self-accelerating
brane-world model \cite{Deffayet:2000uy}. However, this is only a
coincidence, as the detailed dynamics of metric perturbations cannot
be identical. For example, the anisotropic stress always vanishes in
{\it Cuscuton} models, which follows from the use of 3+1D Einstein
equations for linear perturbations of a scalar field. On the contrary,
the anisotropic stress is generically non-vanishing (see e.g.
\cite{Sawicki:2006jj} and references therein) in DGP models.
Nevertheless, this coincidence can be used to examine the observable
differences between dark energy and modified gravity models with exact
same background dynamics.

Before concluding this section, it is important to emphasize that
{\it Cuscuton} is classically a theory of modified gravity, rather
than Einstein gravity with additional scalar field degree of
freedom. This can even be seen at the level of background equations,
as combining Eqs. (\ref{0thfield}-\ref{friedmann}), (for $\dot{\v}
<0$) we can write the modified Friedmann equation as
\begin{equation}
H^{2}=\frac{1}{3M_{p}^{2}}\left\{
\rho_{m}+V\left[V'^{-1}(3\mu^2H)\right]\right\},
\label{eq:modifiedfried}\end{equation} where $V'^{-1}$ is the
inverse function of $V'(\v)$. We see that, in general,  $H^{2}$ is
no longer linearly dependent on the energy density $\rho_{m}$ and
the exact nonlinearity is controlled by the choice of function
$V(\varphi)$.
Moreover, unlike in ordinary Einstein-Hilbert action coupled to a
homogeneous scalar degree of freedom, the modified Friedmann
equation is fixed once $\rho_{m}$ is fixed, i.e. one does not need
to specify initial/boundary conditions for $\varphi$.

To obtain intuition for how the modification works, let us again
consider the quadratic potential. If
$V(\varphi)=\frac{1}{2}m^{2}\varphi^{2}$, it is simple to check that
Eq. (\ref{eq:modifiedfried}) 
is identical to the ordinary Einstein-Hilbert Friedmann equation
with a renormalized Planck's constant \beq M_{p}^{2}\rightarrow
M_{p}^{2}-\frac{3\mu^{4}}{2m^{2}}.\label{planck}\eeq
%
This is a manifestation of {\it Cuscuton} modification of gravity.

This renormalization of the Planck mass is reminiscent of a similar
effect for Lorentz-Violating vector fields
\cite{2004PhRvD..70l3525C}, as well as the exponential quintessence
model (see Sec. \ref{sec_quad}), although, in contrast to {\it
Cuscuton},  both models do introduce an additional dynamical degree
of freedom.

\section{Linear Perturbations in {\it Cuscuton}
cosmology}\label{linear} \subsection{Linearized Field
Equation\label{field_perturbations}}

Varying the {\it Cuscuton} action with respect to $\v$ in a general
curved space-time yields \beq \left(g_{\mu\nu} -
{\partial_{\mu}\v\partial_{\nu}\v\over
X}\right)\nabla^\mu\nabla^\nu\v + \mu^{-2}\sqrt{X}V^{\prime}(\v) =
0, \eeq where $X = \partial^\lambda\v\partial_\lambda\v$, and
$\nabla^\mu$ denotes covariant derivative. Assuming a linearly
perturbed FRW metric in the longitudinal gauge \beq ds^2 =
(1+2\Phi)dt^2-a(t)^2(1-2\Phi)dx^i dx^i, \eeq we can evaluate the
field equation at the linear order in field/metric perturbations:
\beq
3\dot{\v}(\dot{\Phi}+H\Phi)+a^{-2}\nabla^2\delta\v-\mu^{-2}|\dot{\v}|V^{\prime\prime}(\v)\delta\v
=0. \label{1stfield}\eeq Here, $\nabla^2$ is the spatial Laplacian
with respect to comoving coordinates. It is interesting to note that
the linear perturbation equations do not include any second order
time derivative.

Taking the time-derivative of Eq. (\ref{0thfield}), we find \beq
{|\dot{\v}|V^{\prime\prime}(\v)\over\mu^2} = -3\dot{H}. \eeq This
yields the following form for Eq. (\ref{1stfield}) in the Fourier
space: \beq \delta\v =
\frac{3\dot{\v}(\dot{\Phi}+H\Phi)}{\frac{k^2}{a^2}-3\dot{H}},
\label{1stfieldk}\eeq explicitly showing that, as pointed out in
\cite{companion2}, {\it Cuscuton} perturbations simply follow metric
perturbations in a non-local way, and do not introduce any
additional dynamical degree of freedom. In other words, similar to
the homogenous equation (\ref{0thfield}), {\it the field equation
only amounts to a constraint condition that relates metric and field
perturbations}.

\subsection{Linear Einstein Equations: Modified
Gravity\label{gravity}}

The Einstein equations for scalar metric perturbations in the
presence of {\it Cuscuton} as well as ordinary (dust) matter
inhomogeneities are simply written as: \bwt \bea a^{-2} \nabla^2\Phi
= 3H(\dot{\Phi}+H\Phi) + (2M_p^2)^{-1} (\delta\rho_m + V^\prime(\v)
\delta\v),\label{G00}
\\
\dot{\Phi}+H\Phi = (2M_p^2)^{-1}
(\mu^2\delta\v+\rho_m\lambda),\label{G0i}
\\
\ddot{\Phi}+4H\dot{\Phi}+ (2\dot{H}+3H^2)\Phi = (2M^2_p)^{-1}
(\mu^2\sgn(\dot{\v})\dot{\delta\v}-\mu^2|\dot{\v}|\Phi-V^\prime(\v)\delta\v),
\label{Gij}\eea where $\lambda$ is the potential of the matter
peculiar velocity $u^i = a^{-1}\partial\lambda/\partial x^i$.

Transforming to the Fourier space, it is interesting to note that
$\delta\v$ can be eliminated by combining the field equations (Eqs.
\ref{0thfield} and \ref{1stfieldk}) and the $G_{00}$ equation (Eq.
\ref{G00}) to yield a modified law of gravity 
\beq \left(k^2\over a^2\right)\Phi +
\left[3H+{9H(2\dot{H}+3H^2\Omega_m)\over 2
\left(\frac{k^2}{a^2}-3\dot{H}\right)}\right](\dot{\Phi}+H\Phi)+(2M^2_p)^{-1}\delta\rho_m
= 0,\eeq  where $\Omega_m = \rho_m/(3M^2_p H^2)$, is the matter
density in units of the critical density of the Universe.\ewt

Thus we notice that $\delta\v$ completely drops out of the linear
gravity (or Poisson) equation, although it modifies the equation in
a non-local way.  This is another manifestation of {\it Cuscuton}
being {\it a theory of modified gravity}, even though it is a
particular limit of k-essence.

It is important to note that this modification (or screening of
gravity) does not affect Newtonian gravity on small sub-horizon
scales, i.e. as long as $\frac{k^2}{a^2} \gg H^2, \dot{H}$.

We should point out that the only other modification of Einstein
gravity that does not introduce a new degree of freedom (to the best
of our knowledge), known as Modified-Source Gravity
\cite{Carroll:2006jn}, is conformally equivalent to the Einstein
gravity. Moreover, Modified-Source Gravity can be constructed by a
non-linear local modification of the matter Lagrangian, where, in
terms of the modified Lagrangian, (and in contrast to {\it
Cuscuton}) the gravity reduces to Einstein gravity.

\subsection{Linear Einstein Equations: Modified
Dynamics\label{gravity}}

 Let us consider the modified dynamics of the
gravitational potential, $\Phi$, in the presence of {\it Cuscuton}
and pressureless dark matter.
 Plugging Eq. (\ref{1stfieldk}) into Eq. (\ref{Gij}), after
straightforward manipulations, we arrive at \bwt \bea
(1+C_{2})\ddot{\Phi}+(4H+C_{1}+C_{2}H+C_{3})\dot{\Phi}+(3H^{2}+\dot{H}-\frac{3}{2}\Omega_m H^2+C_{1}H+C_{2}\dot{H}+C_{3}H)\Phi=0,\label{eq:finalisweq}\\
C_{1}\equiv\frac{3(\ddot{H}+3H\dot{H})}{\frac{k^2}{a^2}-3\dot{H}},
C_{2}\equiv\frac{3(2\dot{H}+3H^2\Omega_m)}{2\left(\frac{k^2}{a^2}-3\dot{H}\right)},
C_{3}\equiv\frac{3\left[2H(\frac{k}{a})^{2}+3\ddot{H}\right](2\dot{H}+3H^2\Omega_m)}{2\left(\frac{k^2}{a^2}-3\dot{H}\right)^{2}},\eea
\ewt which is the desired differential equation for $\Phi$. Note
that in the limit of pure matter dominated case without {\it
Cuscuton} modification of gravity, we have\beq \Omega_m \rightarrow
1, ~{\rm and}~ 2\dot{H}+3H^2, C_{i}\rightarrow0,\eeq implying that
the solution to Eq. (\ref{eq:finalisweq}) asymptotically approaches
$\Phi\rightarrow$constant. 
Also, note that any nontrivial scale dependence introduced by {\it
Cuscuton} is characterized by the scale $\dot{H}$, and not $H$.

Consider the long wavelength limit $(k/a)^{2}\ll\dot{H}$. First, let
us ask whether a constant $\Phi$ is a solution to linearized
Einstein equations with just pressureless dust field degrees of
freedom.  The {\it Cuscuton} modification that could prevent $\Phi$
from being a constant is the coefficient of $\Phi$ in
Eq.~(\ref{eq:finalisweq}):\begin{eqnarray}
3H^{2}+\dot{H}-\frac{3}{2}\Omega_m H^2+C_{1}H+C_{2}\dot{H}+C_{3}H
\nonumber \\ =
3\Omega_m\left(\frac{H\ddot{H}}{2\dot{H}^{2}}-1\right)H^2,
\end{eqnarray} which generically does not vanish unless the scale
factor is of the form\begin{equation}
a=a_{i}[1+\frac{|\dot{H}_{i}|}{H_{i}}(t-t_{i})]^{H_{i}^{2}/|\dot{H}_{i}|}\label{eq:oneovert},\end{equation}
corresponding to power-law expansion, or a constant effective
equation of state. Therefore, having a constant effective equation
of state will allow a constant $\Phi$ solution on long wavelengths.

 Next, whether or not $\Phi$ is damped (due to the friction term in Eq. \ref{eq:finalisweq})
depends on the sign of \begin{equation}
\frac{4H+C_{1}+C_{2}H+C_{3}}{1+C_{2}}=H+\frac{d}{dt}\ln
\dot{H}^{-1}.\end{equation} Hence, we see that whether the potential
decays or not depends on how fast $\dot{H}$ horizon grows/decays
with time. In the case a constant effective equation of state (or
Eq. \ref{eq:oneovert}), the potential will always decay until it
asymptotically reaches a constant. As we will argue in Section
\ref{sec::causality}, the general behavior of $\Phi$ on superhorizon
scales can be easily understood from the conservation of the Bardeen
parameter.

In the opposite limit of $(k/a)^{2}\gg\dot{H}$, the coefficient in
Eq. (\ref{eq:finalisweq}) that prevents $\Phi$ from being constant
becomes\begin{equation}
3H^{2}\left(1-\frac{\Omega_m}{2}\right)+\dot{H}
\end{equation} which again does
not necessarily vanish with the gravity modified by a nontrivial
$V(\varphi)$, even though the form of the terms looks identical to
that of Einstein gravity sourced by pressureless dust.  That is
because the background Einstein equations are modified by the
presence of {\it Cuscuton}. Note that, as far as the damping
coefficient is concerned, since $C_i \rightarrow 0$ in the short
wavelength limit, the $\dot{\Phi}$ term always acts as a damping
term.

\section{Observational signatures of {\it Cuscuton} cosmologies}
\label{sec::linear} In this section, we study the observational
signatures of  {\it Cuscuton} cosmology.  First we consider the
observables analytically, in a perturbative setting. Afterwards, we
focus on different observational signatures and constraints for
quadratic and exponential {\it Cuscuton} potentials.

\subsection{Analytic Treatment of General {\it Cuscuton} Potentials}

In addition to the matter power spectrum, the decaying gravitational
potential caused by the {\it Cuscuton} modification will manifest
itself in the anisotropy of the Cosmic Microwave Background (CMB).
The induced CMB anisotropy due to the Fourier mode $\Phi_{\bf k}$ is
\cite{2003moco.book.....D}:
\begin{eqnarray}
\Theta_{l, \bf k} & = & \int_{0}^{\eta_{0}}d\eta
~g(\eta)(\Theta_{0}+\Phi_{\bf k})j_{l}[k(\eta_{0}-\eta)]+ \nonumber \\
 &  & \frac{1}{ik}\int_{0}^{\eta_{0}}d\eta~
v_{b}({\bf k}) g(\eta)\frac{\partial}{\partial\eta}j_{l}[k(\eta_{0}-\eta)]+ \nonumber \\
& & 2\int_{0}^{\eta_{0}}d\eta~\frac{\partial\Phi_{\bf
k}}{\partial\eta}e^{-\tau(\eta)}j_{l}[k(\eta_{0}-\eta)],
\label{eq:isweffect}
\end{eqnarray}
where $\eta$ is the conformal time, $\Theta_0$ is the monopole
temperature fluctuation, $g(\eta)$ is the visibility function, $v_b$
is the baryon velocity field, $\tau$ is the optical depth, and
$\eta_0$ is the present day conformal time.  The first term is known
as the Sachs-Wolfe effect, while the second term is the dipole
contribution, and the last term is the Integrated Sachs-Wolfe (ISW)
contribution, which depends on the time variation of the
gravitational potential. All the integrals are taken over the light
cone.

We can compute the ISW contribution to CMB anisotropies perturbatively. 
To first order in the {\it Cuscuton} potential, $V$, one can write
$\partial\Phi/\partial\eta$, which appears in the ISW term of
Eq.~(\ref{eq:isweffect}), as
\begin{equation}
\frac{\partial\Phi_{(1), {\bf k}}}{\partial\eta} = a_i
\left(\frac{t(\eta)}{t_i}\right)^{-2} \int_{t_i}^{t(\eta)} dt'
\left(\frac{t'}{t_i}\right)^{8/3} S_{\bf k}(t'),\label{phidot}
\end{equation}
where $t(\eta)\sim a_i^3 \eta^3/27t_i^2$, $S(t)$ is given by \bwt
\begin{equation}
S_{\bf k}(t)\equiv -\left\{3V+\frac{3t\dot{V}}{4}+
\frac{3}{\frac{k^2}{a^2}+\frac{2}{t^{2}}}\left[\frac{\dot{V}}{t}+\frac{\ddot{V}}{2}+\left(\frac{\dot{V}}{2t}\right)\frac{\frac{4k^2}{3a^2}+\frac{4}{t^{2}}}{\frac{k^2}{a^2}+\frac{2}{t^{2}}}\right]\right\}\frac{\Phi_{(0),
{\bf k}}}{3M^2_p}, \label{eq:pertsource}
\end{equation}
\ewt where $\Phi_{(0), {\bf k}}$ is the zeroth order solution in
$V$, to Eq. (\ref{eq:finalisweq}), which is a constant in a flat
matter dominated universe.  The key feature manifested in
Eq.~(\ref{eq:pertsource}) is the mildness of the scale dependence.
This is simply a result of the fact that $C_i \rightarrow 0$ in the
limit of $k\rightarrow \infty$, while $C_i$ does no vanish in the
limit of $k\rightarrow0$.  Although more explicit expressions for
$\Theta_l$ and $C_l$ may be derived, general results deviating
strongly away from cosmological constant-induced ISW effect are
complicated and unilluminating. Hence, in the subsequent
subsections, we will numerically examine two interesting explicit
potentials, and compare them against current cosmological
observations.

\begin{figure*}
\includegraphics[width=0.48\linewidth]{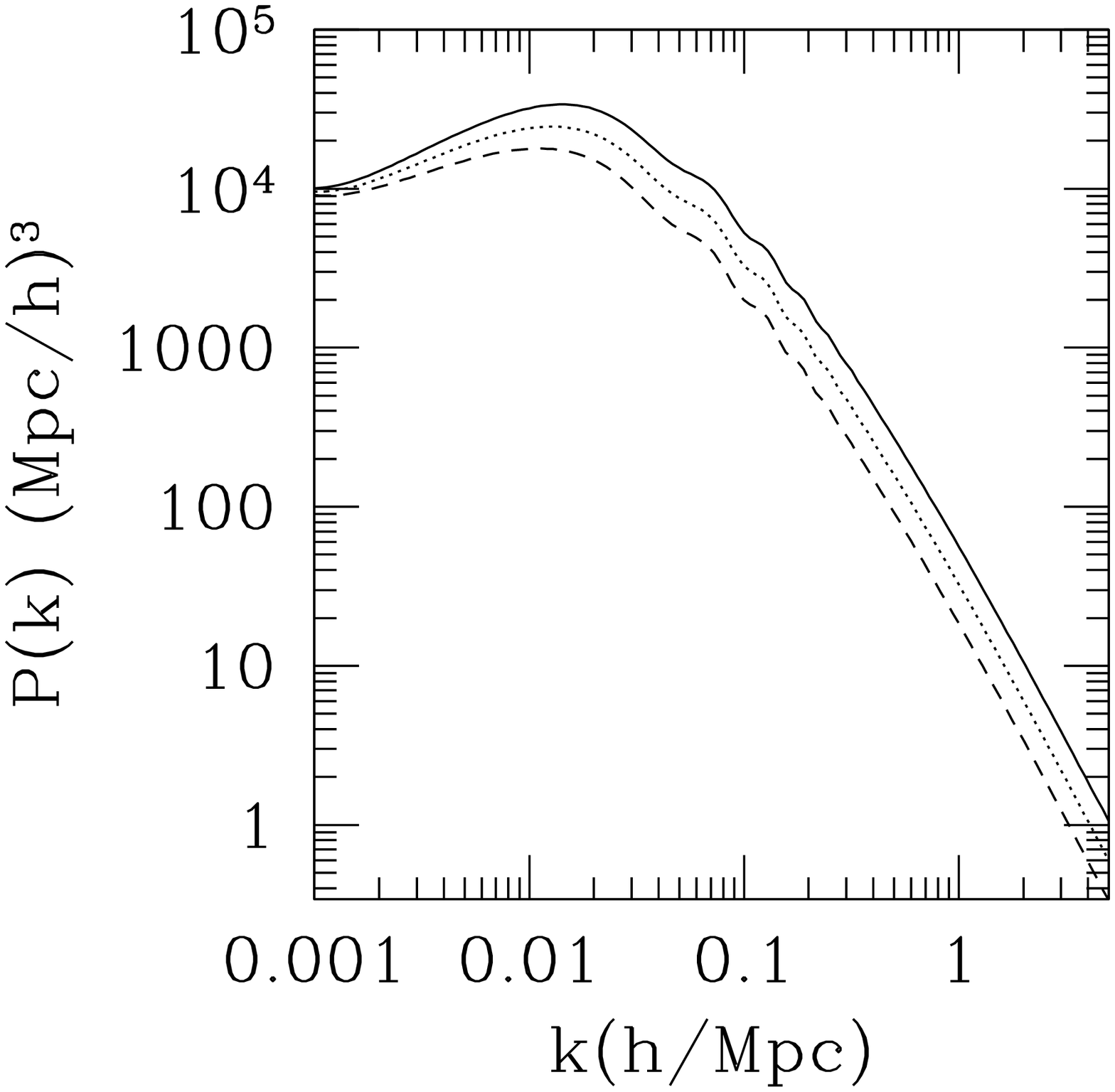}
\includegraphics[width=0.48\linewidth]{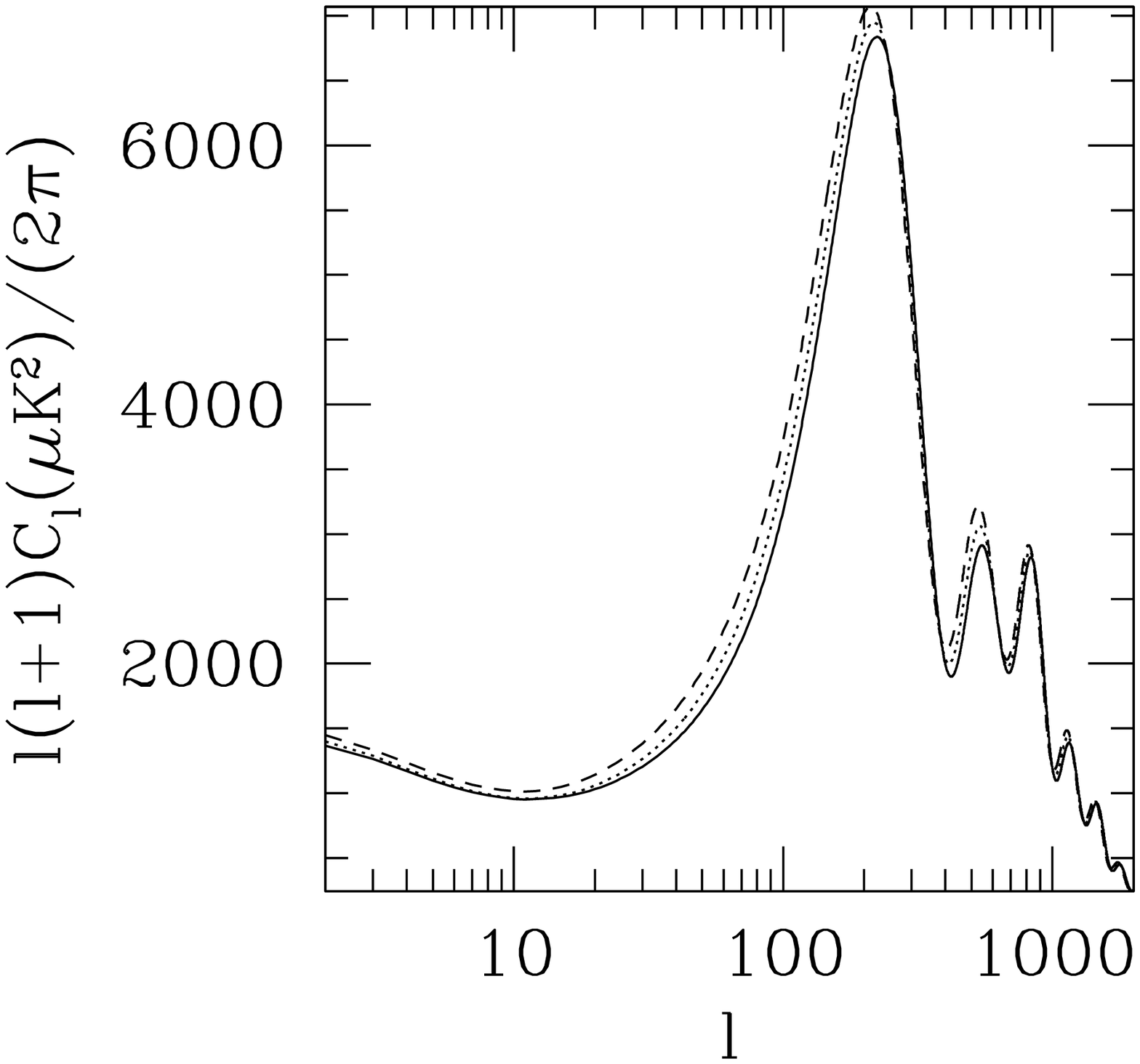}
\caption{\label{pspec} {\it Left}: Dark matter power spectra for
quadratic {\it Cuscuton} densities $\Omega_Q = 0,0.05,0.1$, shown as
solid, dotted, and dashed lines respectively. The initial amplitude
of the scale-invariant adiabatic perturbations, is kept constant.
{\it Right}: The CMB anisotropy power spectrum for the same
cosmologies. }
\end{figure*}

\subsection{Quadratic (Tracking) Potential}\label{sec_quad}
Using potentials of the form $V(\v)= V_0+\frac{1}{2}m^2\v^2$, one
finds solutions similar to that of a tracking dark energy
\cite{Wetterich:1987fm,Ratra:1987rm,Caldwell:1997ii} component plus
a cosmological constant. While $V_0$ simply contributes towards the
cosmological constant, the quadratic term $\frac{1}{2}m^2\v^2$
maintains a constant fraction of the total energy density of the
Universe, much like models of early dark energy
\cite{Hebecker:2000zb,Doran:2006kp,Linder:2006ud}. To see this, take
the square of Eq. (\ref{0thfield}) together with the Friedmann
Equation (\ref{f1}) to get \beq \Omega_Q = -\frac{\Delta
M^2_p}{M^2_p} = \frac{\frac{1}{2}m^2\v^2}{\rho_{tot}} =
\frac{3\mu^4}{2M^2_p m^2} = {\rm const.}, \eeq where $\Delta M^2_p$
denotes the equivalent change in the large scale Planck mass (Eq.
\ref{planck}). Therefore, assuming $V_0 \neq 0$, the expansion
history $H(z)$ is exactly equivalent to that of a $\Lambda$CDM
cosmology, as the extra quadratic component simply follows the rest
of the energy content of the Universe. In fact, the expansion
history $H(z)$ becomes independent of $\Omega_Q$, provided one
scales the other components of the Universe appropriately, i.e.
simultaneously transforming $\Omega_i \rightarrow
(1-\Omega_Q)\Omega_i$, where $i$ labels $\Lambda$, photons, baryons
and dark matter while changing the number of relativistic neutrino
species   $N_{\nu} = 3.04 (1-2.47\Omega_Q)$ leaves $H(z)$ invariant.
This leads to two corollaries: First, there will be a degeneracy
between $N_{\nu}$ and $\Omega_Q$ (at least, as long as we only
consider the expansion history). Secondly, quadratic {\it Cuscuton}
can only be constrained by studying sub-horizon perturbations,
provided one does not a priori know the number of neutrino species.

Current constraints on light element abundances, which are predicted
from Big Bang Nucleosynthesis, already put significant constraints
on the number of relativistic neutrinos during the radiation era,
$N_\nu = 3.1 \pm 0.7$ \cite{2005APh....23..313C}, which yields
$\Omega_Q \lesssim 10\%$. However, as we see below, observational
constraints on cosmological inhomogeneities can put much tighter
constraints on $\Omega_Q$.


Since {\it Cuscuton} perturbations have an infinite speed of sound,
they do not cluster on sub-horizon scales, and thus perturbations
start to decay as they enter the horizon. This behavior is
reminiscent of scalar field dark energy (or quintessence) for which,
once inside horizon, the perturbations also free-stream with the
speed of light.

One can work out this sub-horizon decay analytically during matter
domination where $a \propto t^{2/3}$. Then, ignoring $\delta\v$ in
Eq. (\ref{Gij}) (or using Eq. \ref{eq:finalisweq} with $C_i
\rightarrow 0$), in the $k \gg aH$ regime, we end up with: \beq
\ddot{\Phi}+4H\dot{\Phi}+\frac{3}{2}H^2 \Omega_Q \Phi =0,\eeq which
can be easily solved if $H=\frac{2}{3t}$: \beq \Phi \propto
a^{\alpha}; \alpha =
\frac{5}{4}\left(-1\pm\sqrt{1-\frac{24}{25}\Omega_Q}\right). \eeq
This result is identical to the growth suppression for a
quintessence field with an exponential potential, which has a
similar tracking behavior \footnote{Although, note that $\Omega_Q$
for an exponential quintessence field depends slightly on the
background equations of state (e.g. matter vs. radiation)} (e.g. see
Eq. 12 in \cite{Ferreira:1997au}). For example, for $\Omega_Q \ll
1$, the dominant metric mode decays as $a^{-3\Omega_Q/5}$ during
matter domination, which can also be obtained from Eqs.
(\ref{phidot}-\ref{eq:pertsource}). This introduces a significant
suppression of large scale structure power during the matter era:
\beq\label{eqn::supp} \frac{\delta(\Omega_Q,z=0)}{\delta(0,z=0)}
\simeq z_{eq} ^{-3\Omega_Q/5}   \simeq 0.61^{\Omega_Q/0.1}, \eeq if
we fix the amplitude at matter-radiation equality, assuming that it
happened at $z_{eq} \simeq 3500$. Of course, only modes inside the
horizon at matter-radiation equality will feel this amount of
suppression. Modes entering after equality will suffer less
suppression the later they enter. This leads to a red tilt in the
cold dark matter spectrum up to the scale of matter radiation
equality $k_{eq}$ mimicking a running spectral index
\cite{Caldwell:2003vp}.
 However, all scales $k>k_{eq}$ are suppressed by the same factor
(\ref{eqn::supp}) which distinguishes quadratic {\it Cuscuton} from
a simple running spectral index $\Lambda$CDM model. As the metric
potential decays inside the horizon through the matter era, the CMB
anisotropies receive an extra contribution from the Integrated
Sachs-Wolfe (ISW).  This leads to additional power, typically
shifting the first Doppler peak of the CMB power spectrum towards
slightly lower multipoles.

Fig. (\ref{pspec}) shows the resulting matter and CMB power spectra
for $\Omega_Q =0,0.05,$ and $0.1$, where we have fixed the initial
amplitude of scale-invariant adiabatic perturbations, as well as the
background expansion history, to that of WMAP3 concordance model
\cite{Spergel:2006hy}. As expected, the changes in the amplitude of
the matter power spectrum (left panel in Fig. \ref{pspec}) is most
significant. However, direct measurement of the amplitude of the
power spectrum (e.g. from galaxy surveys) may be problematic due to
the ambiguity in the value of linear bias. Methods to measure the
linear bias, while present, are not as reliable as the shape of the
power spectrum, as they involve non-linear physics, and thus are not
widely used to obtain cosmological constraints. In addition,
analytic calculations of non-linear structure formation in similar
model of early dark energy yield considerable and quite unexpected
deviations from the $\Lambda$CDM scenario precluding the use of
standard approximations to infer the non-linear spectrum given the
linear evolution \cite{Bartelmann:2005fc}.

Thus, instead of galaxy catalogs, we use Lyman-$\alpha$ forest
observations of quasar spectra from the Sloan Digital Sky Survey
(SDSS) \cite{Seljak:2006bg}, which mainly constrains the linear
overdensity $\Delta^2_L$ and spectral index $n_{\rm eff}$ at a pivot
scale of $k \simeq 0.009~ {\rm s/km}$ ($\simeq 1 ~{\rm Mpc}^{-1}$),
and a pivot redshift of $z\simeq 3$ \cite{Seljak:2006bg}. In this
mildly non-linear regime, we anticipate non-linear effects of
$\Omega_Q$ to play little role. This point is further strengthened
from the low values of $\Omega_Q$ allowed by all other data without
using Lyman-$\alpha$ (see below). At such low values of $\Omega_Q$,
the remaining non-linear effects of $\Omega_Q$ will be negligible.
Hence, Lyman-$\alpha$ forest observations seem well suited to
constrain the effects of $\Omega_Q$ on growth of linear
perturbations.

The changes in the CMB power spectrum are more subtle. As mentioned
above, the main impacts of the {\it Cuscuton} quadratic potential
can be seen in an additional ISW contribution and a slight
suppression of small scale power in the CMB spectrum. We ran a Monte
Carlo Markov Chain (MCMC) simulation to constrain this  model using
a modified version of cmbeasy \cite{Doran:2003sy}.  For this, we
used the 3-year CMB data of WMAP \cite{Spergel:2006hy}, Supernovae
Ia measurements \cite{Astier:2005qq,Riess:2004nr}, constraints from
baryonic acoustic oscillations \cite{Eisenstein:2005su}, as well as
the latest release of the SDSS galaxy survey \cite{sdsslrg} and SDSS
Lyman-$\alpha$ forest observations \cite{Seljak:2006bg}.

The WMAP 3-year data alone constrains $\Omega_Q < 3.6\%$ at 95\%
confidence level. Adding SNe Ia and large scale structure data
improves this -- already substantial -- bound further to  $\Omega_Q
< 2.7\%$. Finally, adding Lyman-$\alpha$ forest data yields a tight
limit of  $\Omega_Q < 1.6\%$.

\begin{figure}
\includegraphics[angle=-90,width=\linewidth]{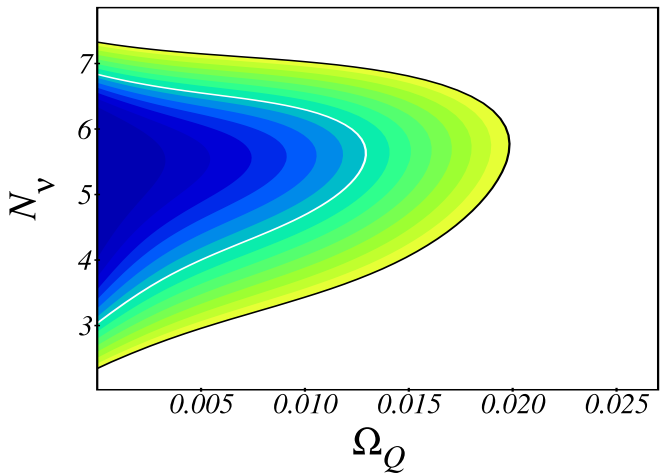}
\caption{\label{fig::contour} Confidence contours for the abundance
of the quadratic {\it Cuscuton}, $\Omega_Q$ and the number of
relativistic neutrino species $N_{\nu}$ derived from using SNe Ia,
CMB, large scale structure and Lyman-$\alpha$ forest observations.
The white and black boundaries indicate 68\% and 95\% likelihood
regions. While in principle, a degeneracy of the background
evolution between $\Omega_Q$ and $N_{\nu}$ exists, it is too weak to
be seen here, because the allowed abundance of $\Omega_Q$ is too
small to have any significant effect on $N_{\nu}$ constraints (see
the text). 
}
\end{figure}

Hence, the background degeneracy of $N_{\nu}$ and $\Omega_Q$ while
present has little effect on the contours in Fig.
(\ref{fig::contour}).   The allowed abundance $\Omega_Q$ is very
small such that the effect on $N_{\nu}$ is at most $\Delta N_{\nu} =
-3.04 \times 2.47 \times \Omega_Q \approx -0.2$. This change in
$N_{\nu}$ is rather small compared to the 68\% confidence interval
for $N_{\nu}$, which is $N_{\nu} = 5.5^{+0.9}_{-1.2}$. 

\subsection{Exponential (DGP-like) Potential}

\begin{figure}
\includegraphics[width=\linewidth]{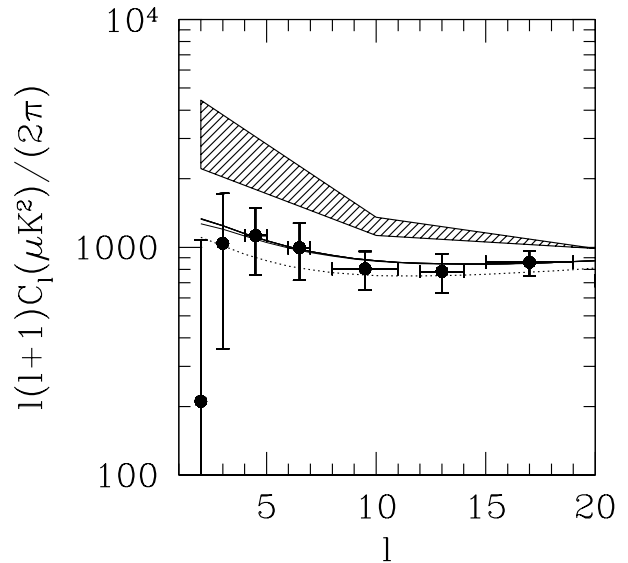}
\caption{\label{dgp_cmb} CMB anisotropy power spectrum at low l's:
The solid curves show the CMB power spectrum for the DGP-like dark
energy models with $c^2_s =1, 10, 100$, where larger sound speeds
have slightly larger power. Exponential {\it Cuscuton} is realized
in the $c^2_s \rightarrow \infty$ limit, although the power spectra
for $c^2_s > 10$ are virtually indistinguishable. The dotted curve
is the power spectrum for the $\Lambda$CDM cosmology, while the
dashed region shows the prediction for the DGP self-accelerating
model \cite{Song:2006jk}. All the power spectra assume
$\Omega_m=0.26$ and $h=0.66$. The points show the WMAP 3yr
observations, while the errorbars indicate an estimate of the cosmic
variance error based on the data \cite{Hinshaw:2006ia}.}
\end{figure}

DGP model of modified gravity \cite{Dvali:2000hr}, similar to other
brane-world models, is inspired by non-perturbative objects in
string theory. It posits that the observable universe lives on a
3+1D brane in 4+1D space-time. While matter fields are constrained
to the brane, gravity can propagate into the bulk and thus can be
sensitive to the full geometry of the Universe. The characteristic
of the DGP model is that the brane action includes a term
proportional to the volume integral of the induced Ricci scalar on
our 3-brane. This is added to the ordinary Einstein-Hilbert action
in the bulk, which integrates the full Ricci scalar. It turns out
that at high energies, the induced action dominates the dynamics of
the induced metric, leading to ordinary four dimensional gravity,
while at low energies, the bulk action takes over and gravity
becomes five dimensional. This may lead to a self-accelerating phase
of cosmic evolution, even in the absence of a cosmological constant.

Although it is argued that the self-accelerating phase of the DGP
model contains ghosts \cite{2005PhRvD..72l3511K,
2006PhRvD..73d4016G, Charmousis:2006pn}, and thus may not be
realized in a physically stable way (but see
\cite{Deffayet:2006wp}), it still remains the most widely studied
concrete example of a modified gravity model which competes with
dark energy models as an explanation for the observed acceleration
of the Universe.

As we mentioned in Section \ref{back}, for a spatially flat 3-brane,
the background evolution within self-accelerating DGP model
coincides with that of a {\it Cuscuton} model with an exponential
potential (Eq. \ref{dgph}). Therefore, it is not possible to
distinguish the two models based on the traditional geometrical
tests of background cosmology, such as supernovae Ia
\cite{Riess,Perlmutter:1998np}, distance to the last scattering
surface, or the scale of baryonic acoustic oscillations
\cite{Eisenstein:2005su}.

For sub-horizon perturbations, \cite{2006JCAP...01..016K} and
\cite{Song:2006jk} compare the linear growth of structure in DGP
model with a quintessence model with an identical expansion history.
Since dark energy does not cluster on sub-horizon scales for $c_s
\geq 1$, such a quintessence model should have a very similar
behavior to the exponential {\it Cuscuton} model. \cite{Song:2006jk}
finds that the DGP model shows an extra 5-10\% decay compared to the
quintessence model (see their Fig. 2). Therefore, it is not easy to
distinguish DGP and a DGP-like {\it Cuscuton} (or quintessence)
model based on the growth of the large scale structure.
\cite{Huterer:2006mv} estimate that this distinction can only be
done at less $3\sigma$ level, with the next generation of
supernovae, weak lensing, and (small angle) CMB observations.

More significant is the ISW effect induced by the anisotropic stress
in the DGP model, which is a characteristic feature of modified
gravity models. \cite{Song:2006jk} find that, as a result of the ISW
effect, the CMB anisotropy power spectrum is a factor of $\sim 4$
larger than than $\Lambda$CDM cosmology on small $\ell$'s
\footnote{This already puts DGP model at odds with WMAP power
spectrum \cite{Hinshaw:2006ia} in the $\ell < 10$ range (see Fig.
\ref{dgp_cmb}). However, the ansatz used in \cite{Song:2006jk} to
calculate metric perturbations may not be applicable on such large
scales.}. Therefore, they advocate use of correlations between CMB
and high redshift large scale structure surveys (see e.g.
\cite{Afshordi:2004kz}) in order to detect signatures of the DGP
model. As we show in Fig. (\ref{dgp_cmb}), the CMB power spectrum
for the exponential {\it Cuscuton} model is nearly identical to that
of the DGP-like quintessence model, and is still much smaller the
prediction for the actual DGP model at low $\ell$'s \footnote{As the
conventional formulation of dark energy perturbations (used in
publicly available CMB Boltzmann codes) uses a finite speed of
sound, we approximate the exponential {\it Cuscuton} model by a dark
energy model with a finite but large speed of sound, but the same
expansion history. However, as is evident from Fig. (\ref{dgp_cmb}),
the power spectra are indistinguishable for $c^2_s
>10$.}.

We thus conclude this section by pointing out that observational
distinction between dark energy and modified gravity models may be
significantly more difficult than advocated in the literature (e.g.
\cite{Ishak:2005zs} where the authors also point to a number of
issues that need to be explored in such endeavor). A relatively
simple and well-behaved dark energy model such as exponential {\it
Cuscuton} can exactly replicate the expansion history of the DGP
self-accelerating cosmology, while predicting similar small angle
CMB and matter power spectra, at a few percent level. The smoking
gun for the modified gravity models, thus, is their anisotropic
stress, that can be potentially probed by the ISW effect in the low
$\ell$ regime of the CMB power spectrum.

\section{On the Causality of {\it Cuscuton} Linear Perturbations}
\label{sec::causality}

In this section, we briefly comment on causal properties of the {\it
Cuscuton} field theory with a minimal gravitational coupling.

Eq. (\ref{1stfieldk}), which simply has a Yukawa screening form, can
be written in the real space as: \beq \delta\v({\bf x},t) =
3\dot{\v}a^2 \int d^3{\bf x'} \frac{e^{-a|3\dot{H}|^{1/2}|{\bf x -
x'}|}}{4\pi |{\bf x - x'}|}[\dot{\Phi}({\bf x'},t)+H\Phi({\bf
x'},t)]. \eeq Therefore, we see that the non-local dependence of
{\it Cuscuton} on the metric perturbations is exponentially
suppressed beyond a horizon that is defined by $|\dot{H}|$. It is
interesting to notice that an $\dot{H}$ horizon, also naturally
occurs in cosmological gravitomagnetism, beyond which, the coupling
of a gyroscope precession to the rotation of its surrounding
environment saturates \cite{Schmid:2005pv}.

Since metric scalar perturbations do not propagate, and tensor
perturbations that do propagate, do not couple to {\it Cuscuton}
perturbations at the linear order, it is not possible to rigorously
address the question of causality of {\it Cuscuton} coupled to
gravity, in the context of what we have done thus far. However, it
is clear that the naive picture of instantaneous interaction, due to
the infinite sound speed of {\it Cuscuton}, does not hold in a
general relativistic context, as {\it Cuscuton} is exponentially
insensitive to metric fluctuations at separations larger than an
$\dot{H}$ horizon. By the same token, and direct substitution into
Eq. (\ref{eq:finalisweq}), one can see that super-horizon
perturbations in the Bardeen parameter \cite{1980PhRvD..22.1882B}:
\bea \zeta = \Phi -\frac{H}{\dot{H}}(\dot{\Phi}+H\Phi)\nonumber\\ =
\Phi +\frac{2(H^{-1}\dot{\Phi}+\Phi)}{3(1+w)},\label{bardeen} \eea
remain constant until the modes enter the $\dot{H}$ horizon, where
$w$ is the effective equation of state. Therefore, {\it there is no
gross violation of causality} in {\it Cuscuton} cosmology.
\section{Conclusions}
\label{sec::conclusions}

In this paper, we have studied the physical features, as well as
possible observational signatures of a cosmology with {\it Cuscuton}
dark energy, which was first introduced in a companion paper
\cite{companion2}, and could be realized as an incompressible
k-essence fluid. We showed that {\it Cuscuton} perturbations have no
independent dynamical degree of freedom and, in lieu of other
couplings, simply follow the space-time metric.  Therefore, {\it
Cuscuton} can be considered to be a minimal theory of evolving dark
energy, or a minimal modification of a cosmological constant. Due to
lack of internal dynamics, {\it Cuscuton} only modifies (or dresses)
the gravity of massive objects, and thus resembles a modified
gravity theory. Indeed, to the best of our knowledge, this is the
only modification of Einstein gravity that does not introduce any
additional degree of freedom (and is not conformally equivalent to
Einstein gravity). We then studied two specific {\it Cuscuton}
cosmologies, with quadratic and exponential potentials.

We saw that the expansion history of {\it Cuscuton} cosmology with a
quadratic potential is identical to that of $\Lambda$CDM, and thus
geometrical tests such as supernovae Ia, or the angular scale of
barynoic acoustic oscillations, are blind to a quadratic term in the
{\it Cuscuton} potential. Nevertheless, the quadratic term acts as
an early dark energy component with a constant energy fraction, and
thus can be detected via its influence on the matter power spectrum,
or through the ISW effect in the CMB. We find that joint constraints
from supernovae Ia, CMB anisotropies, power spectra of galaxy
surveys, and Lyman-$\alpha$ forest fluctuations limit this component
to $\Omega_Q < 1.6\%$, which is the most stringent upper limit that
has ever been put on an early dark energy component.

For the exponential {\it Cuscuton} model, we found that,
surprisingly, the expansion history is identical to that of a flat
DGP self-accelerating modified gravity model. The only observational
distinction between the two cosmologies is in the ISW effect at
$\ell \lesssim 20$ in the CMB power spectrum, and is due to the
anisotropic stress, present in the DGP modified gravity model.

Indeed, cosmologists have yet to develop efficient techniques to
detect a smoking gun for modified gravity. Exponential {\it
Cuscuton} cosmology is a clear example of where such a smoking gun
may become necessary, and detection of a non-vanishing anisotropic
stress (see e.g. \cite{2006ApJ...648..797B}), through ISW effect,
appears to be the only way to rule out any such (scalar) dark energy
model. An independent motivation for why new physics might be at
work in the generation of large angle ISW effect may also come from
observations of large angle anomalies in the CMB sky (e.g.
\cite{Copi:2006tu}, and references therein).

Finally, we showed that, despite its infinite speed of sound, {\it
Cuscuton} perturbations are exponentially insensitive to the metric
perturbations beyond the Hubble radius, justifying why there is no
gross violation of causality for superhorizon perturbations.

It is now becoming clear that the phenomenology of dark energy will
be the central theme in theoretical and observational cosmology,
over the next ten years \cite{Albrecht:2006um}. Nevertheless, all
the current theoretical models for any observable deviation from a
cosmological constant are at the best ill-motivated, or at the
worst, already ruled out. Although falling short of solving the
celebrated ``cosmological constant problem" \cite{ccproblem}, {\it
Cuscuton} dark energy provides an alternative which could be
considered more natural, as it is protected against quantum
corrections \cite{companion2} (even in the presence of other
couplings), and yet allows for a potentially observable evolving
dark energy. The enthusiastic cosmologist may thus find some use for
{\it Cuscuton} in trying to convince the skeptical theorist about
the benefits of cosmological dark energy probes!

\begin{acknowledgments}
We would like to thank Wayne Hu, Dargan Huterer, Christoph Schmid,
and Paul Steinhardt for useful discussions. The work of DJHC and GG
was supported by DOE Outstanding Junior Investigator Program through
Grant No. DE-FG02-95ER40896 and NSF Grant No. PHY-0506002.  NA
wishes to thank the hospitality of the Physics department at the
University of Wisconsin-Madison throughout the course of this
project, as well as the
\href{http://cosmocoffee.info/}{cosmocoffee.com}\footnote{${\rm
\href{http://cosmocoffee.info/}{http://cosmocoffee.info/}}$} weblog,
where this project was originated.
\end{acknowledgments}

\bibliographystyle{utphys_na}
\bibliography{cus_cosmo6}

\end{document}